# Why the hyperbolic polaritons are hyperbolic?

Xiaoyu Xiong[1,†], Le Zhou[1,†], Yihang Fan[1], Weipeng Wang[1], Yongzheng Wen[1], Yang Shen[1,*], Zhengjun Zhang[1,*], Jingbo Sun[1,*], Ji Zhou[1,*]

**Affiliations**

*Corresponding author. Email: shyang_mse@mail.tsinghua.edu.cn,

zjzhang@tsinghua.edu.cn, jingbosun@tsinghua.edu.cn, zhouji@tsinghua.edu.cn

†These authors contributed equally to this work.

**Abstract**

Polaritons travelling along a hyperbolic medium's surface have recently sparked significant interest in nanophotonics for the unprecedented manipulation ability on light at the nanoscale in a planar way, promising potential nano-optical applications, especially in two-dimensional circuitry. Despite of being named hyperbolic polaritons, the hyperbolic nature has not been thoroughly revealed since an analytical description of the Iso-frequency contour is still elusive. In this work, we proposed an analytical form for describing the iso-frequency contour of the hyperbolic polaritons, showcasing their strictly hyperbolic nature. Such an analytical form is obtained based on the focusing behavior of the hyperbolic polaritons and verified by both the published data from commonly used hyperbolic media systems of the hyperbolic polaritons and our own experimental characterizations on a hyperbolic metamaterial film. By presenting a concise and intuitive physical image, this work may provide a groundbreaking methodology in developing novel hyperbolic polaritons based optical devices.

## Introduction

When materials' anisotropy becomes hyperbolic, light can be manipulated below the diffraction limit. Materials with hyperbolic dielectric tensors have been widely used in realizing a subwavelength focusing in both the bulky way, like the hyperlens [1-5] and the in-plane design [6 -10] with the interface. The sustained mode of the light could be excited phonon polaritons but also the Dyakonov Surface waves [11-13], covering the wavelength range from Mid-Infrared to visible range, obtained along the interface of the hyperbolic media including Van der Waals materials (vdWms), such as alpha-$MoO_3$ [14-23], $V_2O_5$ [24], $MoOCl_2$ [25] patterned hexagonal boron nitride (h-BN) [26-33], bulky crystal of strong anisotropy, e.g. calcite [34], as well as the hyperbolic metamaterials (HMMs) [35-40]. Due to the extremely strong confining effect, light can be trapped to the deep subwavelength scale, which allows potential applications in nonlinear optics [41, 42], subwavelength imaging and focusing [43-46], optical vortex generation [47]and biosensing [48, 49] by enhancing the light-matter interactions. Similar to the bulky materials case, the ability to realize such an ultra-strong confinement under the diffraction limit can also be fundamentally attributed to the hyperbolic anisotropy, which exhibits a hyperbola like Iso-Frequency Contour (IFC) along the plane of the interface. Despite many pioneering experimental studies [11-31] showing the focusing behavior along the interfaces of various hyperbolic materials, a clear theoretical demonstration that the IFC is analytically hyperbolic is still missing. The challenge exists in the complicated system that usually involves in three different materials with the one in between highly anisotropic. Thus, it is very difficult to obtain a precise analytical form, just like the quadratic function in bulky hyperbolic materials case [1-5, 33] that can describe the IFC without any approximations. So far, the widely used IFCs (in methods) were originated from a thin slab model (TSM), expressed by such a complicated equation [50] that the IFC had to be plotted through numerical calculations. The plotted curve showed a hyperbolic like shape, which was then considered as a hyperbola, with its asymptotes going across the zero point in the *k* space coordinate. Such an IFC was always used to demonstrate the ultra-tight focusing, yet cannot be proved as a hyperbola and hardly show the hyperbolic nature directly [50]. Here, by investigating the focusing behavior of the hyperbolic polaritons, we theoretically obtain the analytical description of the IFC, which is proved to be indeed a hyperbola. Basic parameters including the hyperbolic medium film's dielectric properties and thickness are considered in this IFC. The obtained hyperbolic equation also reveals that the center of the hyperbolic IFC is not at the zero point one while with a certain shift towards wave vector along the negative permittivity in the $k$ space for the first time. Furthermore, the validity of the IFC equation is tested by the data achieved from actual materials' parameters, including α-MoO$_3$, patterned h-BN, and a HMM film made of silver grooves structure, and the results are compared with the IFCs obtained by traditional methods, including the equation from the TSM and the numerical fitting with the Fourier transformation (FT) based on the E field distribution [17, 18, 32]. Finally, we perform the experimental characterization on the hyperbolic polaritons along the HMM film by using a scattering-type near-field scanning optical microscopy (s-SNOM) in transmission mode, which further verifies the IFC through the fitting of the measured amplitude and equi-phase contour. This analytical description of the hyperbolic IFC reveals the fundamental mechanism of the anisotropic polaritons' propagation behavior, which may fertilize the new designs of the in-plane photonic devices based on the hyperbolic media.

**Results**

**The general analytical dispersion of surface polaritons**

The IFC is actually describing angular dispersion behavior of the light beam's wave vector ($k$) at the certain wavelength inside a material or along the material's interface if the light beam is in a surface wave mode, which is the case of the hyperbolic polaritons studied in this work. If we use a light source that can supply an omnidirectional generation of the surface waves along the interface, an envelope of the electric field would be presented due to the interface's property. For instance, the surface plasmon polariton (SPP) can be generated from a round nano-hole with a radius $R$ in a metal film when a light beam is incident onto the hole, as shown in Fig. 1A. Since the metallic surface is isotropic, $k_\theta$ and Poynting vector $S$ are collinear. The wave vectors of the generated SPPs are perpendicular to the round hole's edge and thus propagating to all directions with the equi-phase plane in a round circle, as shown in Fig. 1B. If the direction of the wave vector $k_{spp}$ is defined by the angle $\theta$ with respect to the x-axis, one can obtain the relation according to the equi-phase:

$$\frac{|k_\theta|}{|k_y|} = 1, \qquad (1)$$

which tells us firstly the refractive index is isotropic in *k* space and secondly the angle band of the SPP is a full band, from 0 to $2\pi$. In this case, the envelop of the E field is infinite, without any forbidden region in the xy plane, as illustrated in Fig. 1C. The angular dispersion of *k* expressed by Eq. 1 is only determined by the materials that compose the interface, no matter what shape of the source used to generate the SPP.

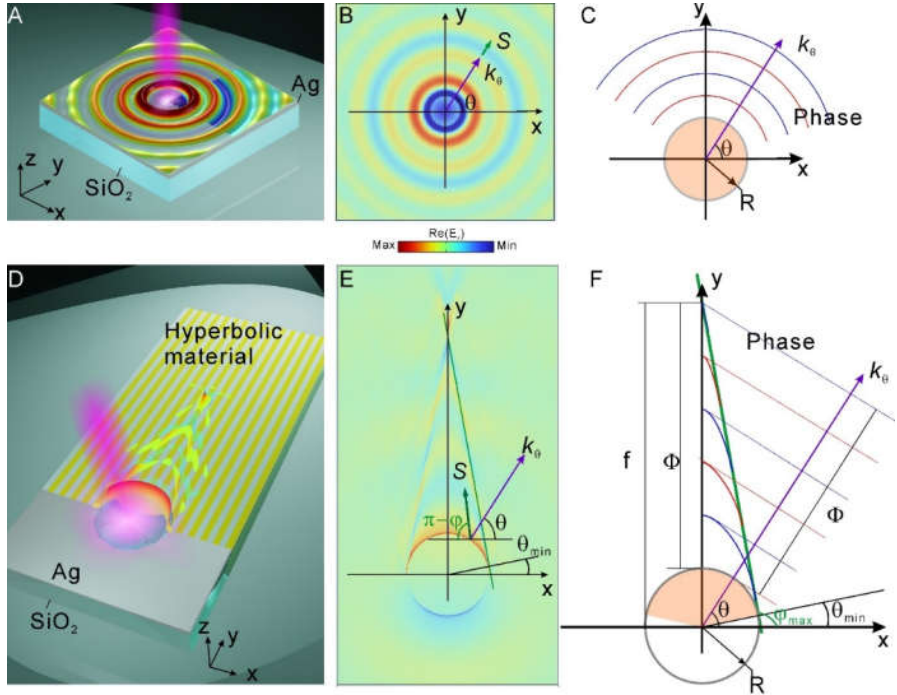

**Fig. 1. The propagation behaviors of the polaritons along isotropic and anisotropic interfaces.** (**A**) SPP generated by a round hole at a silver film on a silica substrate. (**B**) Numerical simulation of the E field of the SPP at the air/silver interface. The direction of the wave vector $k_\theta$ is defined by the angle $\theta$ with respect to the x-axis. $k_\theta$ and Poynting vector *S* are collinear; (**C**) Isotropic phase: $k_\theta$ along all directions has the same amplitude; (**D**) Hyperbolic polariton generated by round hole at a hyperbolic material film with $\varepsilon_x > 0 > \varepsilon_y$; (**E**) Numerical simulation of the E field of the hyperbolic polaritons at the air/hyperbolic material interface, which shows a triangular envelope. $k_\theta$ can only exist in the angle band of $\theta_{min} \leq \theta \leq \pi/2$. The direction of the Poynting vector $S$ is defined by the angle $\varphi$ with respect to the x-axis. Therefore, the Poynting vector at $\varphi_{max}$ define the boundary of the triangular envelope, as illustrated by the green line. (**F**) Anisotropic phase within an angle band of $\theta_{min} \leq \theta \leq \pi/2$. After propagation the phase of $\Phi$, all the generated polaritons meet together at the tip, finally forming a focusing behavior. The structure of the HMM and settings in the numerical simulation can be found in Supplementary Fig. S1. (**A**) and (**C**) share the same color bar.

Now, let us consider the case of the hyperbolic polariton, also generated by a round hole in a hyperbolic medium film with an in-plane anisotropy $\varepsilon_x > 0 > \varepsilon_y$, as shown in Fig.1D. Because of the anisotropy, the equi-phase plane is seriously bent and the Poynting vector $S$ is always not collinear with the wave vector $k_\theta$, as shown in Fig. 1E. Here we use angle $\varphi$ to define the direction of the Poynting vector with respect to the x-axis. Also, the angle band of the wave vector is squeezed to $\theta_{min} \leq \theta \leq \pi/2$, which in turn, limits the angle range of the Poynting vector to be $\pi/2 \leq \varphi \leq \varphi_{max}$. One should notice that $\varphi_{max} - \theta_{min} = \pi/2$ in geometry, while $\varphi_{min} = \theta_{max} = \pi/2$, where the polariton is propagating along the y direction. Here, the $\theta_{min}$ is usually defined as $|\tan\theta| = \sqrt{-\frac{\varepsilon_x}{\varepsilon_y}}$ [50]. Consequently, only those polaritons whose wave vectors fell in this allowed angle band can be generated and propagate along the hyperbolic interface, which forms a triangle envelop of the E field, as shown in Fig. 1E and F. All the energy flow is funneling inside this angular range of $2\theta_{min}$, moving towards the tip of the triangle. Outside of this region, surface mode is forbidden. After propagating the phase of $\Phi$, all polaritons of different $k$ meet together, which is actually the origin of the ultra-tight focusing behavior. In such a case, we can still obtain the relation of wave vectors along different directions:

$$f\sin\theta - R = \frac{\Phi}{k_\theta} \qquad (2a)$$

$$f - R = \frac{\Phi}{k_y} = \frac{\Phi}{k_{min}} \quad (2b)$$

where the wave vector $k_y$ is along y axis and reaches to its minimum $k_{min}$. So, $k_{min}$ is the wave vector with $\theta = \pi/2$, and $k_\theta$ is the wave vector at an angle of $\theta$. $k_\theta$ reaches its maximum $k_{max}$ when the wave vector is at $\theta = \theta_{min}$. The ratio between $k_{min}/k_\theta$ is defined as the normalize refractive index $N$ of the hyperbolic polariton, and it can be obtained with Eq. 2b divided by Eq. 2a:

$$\frac{f - R}{f \sin\theta - R} = N. \quad (3)$$

Thus, the wave vector along a certain direction can be written as a function of $\theta$:

$$\frac{f/R - 1}{f/R \sin\theta - 1} k_{min} = k_\theta, \quad (4)$$

which is the right form of a hyperbola in a polar coordinate. The value of $p = (1 - 1/e) k_{min}$ is usually called directrix of the hyperbola, and the value $e = f/R$ is called eccentricity. To make it even more understandable, Eq. 4 is transformed into its form in a Cartesian coordinate:

$$\frac{(k_y - c)^2}{a^2} - \frac{k_x^2}{b^2} = 1, \quad (5)$$

where $a = (f/R + 1)^{-1} k_{min}$, $b = \sqrt{\frac{f/R - 1}{f/R + 1}} k_{min}$, and $c = \frac{f/R}{f/R + 1} k_{min}$.

Before, lacking an analytical interpretation, the plotted IFCs based on TSM or FT of the E field image showed a hyperbola like curves, and were thought to be located with its center at zero point in the $k$ space, as used in many literatures [14-18, 20]. Unlike the previous illusion, the analytical expression Eq. 5 proves that the shape of the IFC is indeed hyperbolic but with a shift $c$ along the $k_y$-axis, as shown in Fig. 2A. It results in the vertex of this hyperbola at the $k_y$-axis equal to $a + c = k_{min}$, which is the minimum wave vector. The asymptotes of the IFC, which also delineate the minimum end of the angle band $\theta_{min} = \text{atan}(\frac{a}{b}) = \text{atan}\left[\frac{1}{\sqrt{\left(f/R\right)^2 - 1}}\right]$, are intersected at $k_y=c$ but not the zero point of the $k$ space, as implied by the dashed line in Fig. 2A. The connection between the IFC and the actual material's dielectric tensor can be built through $\theta_{min} = \text{atan}\left(\sqrt{-\frac{\varepsilon_x}{\varepsilon_y}}\right)$, and thus, the key parameters $a$, $b$, and $c$ can be fully expressed with $\varepsilon_x$, $\varepsilon_y$, and $k_{min}$. It tells us that the general profile of the hyperbolic IFC is mainly determined by the in-plane anisotropy ($\varepsilon_x \neq \varepsilon_y$), while the thickness of the hyperbolic medium film and the out-plane permittivity ($\varepsilon_z$) may affect the location and the shape of the IFC through $k_{min}$. The relations among all the geometric parameters of the IFC are also shown in Fig. 2A. One should also note that the chosen round shape source is to supply an isotropic generation and then let angular dispersion show up under the hyperbolic dielectric property. Therefore, the IFC of Eq. 5 is the inherent nature of the hyperbolic polaritons along the hyperbolic media's surfaces and is applicable to hyperbolic polaritons generated by any type of sources. Moreover, Eq. 5 describes the IFC in a half space with $k_y>0$. In the other half space of $k_y<0$, the dispersion behavior would be the same according to the inherent symmetry of the materials and thus the IFC is also a hyperbola while shifted to $-k_y$ with $-c$. The total expression of the IFC can be found in Supplementary Fig. S2.

$k_{min}$ is the only unknown parameter in the analytical IFC (Eq. 5). It is the wave vector when the beam is propagating along the y direction of the surface, which can be measured directly either in an experiment or a numerical simulation, or precisely calculated from the eigen-equation of the TM mode in a slab waveguide [51]. In a particular case, if the hyperbolic media is a uniaxially anisotropic with $\varepsilon_y=\varepsilon_z<0<\varepsilon_x$, $k_{min}$ is the wave vector of the SPP propagation mode in a metal core (with permittivity of $\varepsilon_y$) waveguide, which is used to map the IFC of the hyperbolic polariton along a HMM film later. Details can be found in Supplementary Fig. S3.

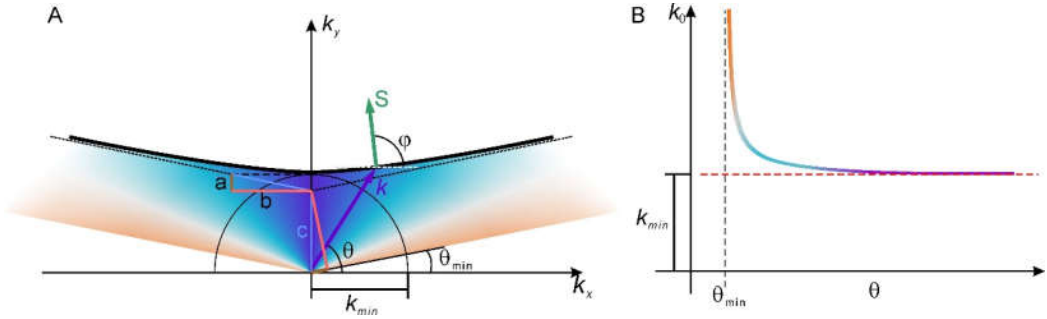

**Fig. 2. Angular dispersion of the anisotropic polariton along a hyperbolic medium interface.**

(**A**) IFC is a hyperbola with a shift of $c$ towards $k_y$ from the zero point. The vertex is located at $(0, a + c)$ and the co-vertex is located at $(b, 0)$. The asymptote is determined by the slope angle of $\theta_{min}$; (**B**) The angularly dispersed wave vector $k$ vs $\theta$, where $k = N \cdot k_{min}$. At the minimum end $\theta_{min}$, $k$ is approaching to infinity, indicated by the color from purple to orange in both Cartesian and polar coordinates.

Another characteristic principle that we can obtain from the IFC is the angular dispersion relation of the refractive index, which can be directly plotted according to Eq. 3. As shown in Fig. 2B, $N$ is monotonically increasing with the decrease of $\theta$. Closing to $\theta_{min}$, the normalized refractive index $N$ and thus $k_\theta = N \cdot k_{min}$ are approaching to infinity, as indicated by the color from purple to orange, which fundamentally ensures the ultra-high confinement of the hyperbolic polaritons along the hyperbolic interface.

*Validity test in different materials system*

To test the validity of this analytical IFC, we applied Eq. 5 to commonly used vdWm, vdWm based metamaterials in Mid-infrared range, as well as the metal based metamaterials in near infrared range. IFC curves calculated by the widely used TSM and fitted from FT of the E field distribution are used as the references to investigate the factors, e.g. thickness and the anisotropy level of the hyperbolic media that may affect the accuracy of the IFC based on different methods. Here we performed numerical simulations of the hyperbolic polaritons generated by a round hole on a α-MoO$_3$ flake [17], patterned h-BN [32] and HMM film of Ag grooves structure [38].

Table I Parameters used in three materials' systems

| Materials | α-MoO$_3$[17] | Patterned h-BN[32] | HMM[38] |
|---|---|---|---|
| Frequency (Wavelength) | 910 cm$^{-1}$ (11989 nm) | 1425 cm$^{-1}$ (7017 nm) | 193.5 THz (1550 nm) |
| Permittivity | $\varepsilon_x$=1.02+0.028i; $\varepsilon_y$=-2.99+0.16i; $\varepsilon_z$=4.81+0.049i; | $\varepsilon_x$=3.7; $\varepsilon_y$=-15.2+0.6i; $\varepsilon_z$=2.1; | $\varepsilon_x$=2.01; $\varepsilon_y$=-64.08+1.64i; $\varepsilon_z$=-64.08+1.64i; |
| $\theta_{min}$(°) | 30.1 | 26.3 | 10.2 |
| Thickness (nm) | 262 | 20 | 35 |
| Substrate permittivity ($\varepsilon_s$) | 5.1+0.18i | 1.07+0.12i | 2.09 |
| $k_{min}/k_0$ | 9.7 | 7.62 | 1.48 |

α-MoO$_3$ is a natural anisotropic material, and may show hyperbolic anisotropy Reststrahlen bands in the Mid-infrared spectral region, supporting the hyperbolic phonon polaritons. Here we choose the frequency of 910 cm$^{-1}$, where [001]: $\varepsilon_x > 0$, [100]: $\varepsilon_y < 0$ and [010]: $\varepsilon_z > 0$ [17]. h-BN is a uniaxial anisotropic material with its optical axis normal to the substrate. Therefore, hBN was patterned into a grating structure to bring in the in-plane hyperbolic anisotropy: $\varepsilon_x > 0$, $\varepsilon_y < 0$, $\varepsilon_z > 0$ [32]. Besides of those hyperbolic media in Mid-infrared, we also consider the HMM film based on silver grooves structure, which shows uniaxial hyperbolic anisotropy ($\varepsilon_y = \varepsilon_z < 0 < \varepsilon_x$) in near-infrared to visible range, as we studied in our former work [38]. The bulky form of the HMM is the Ag/Air multi-layers vertical to the substrate. Once the thickness of the HMM decreases to the level of the period or even less, the vertical layers degrades to grooves. The hyperbolic permittivity is calculated by using Maxwell−Garnett theory. Details can be found in the supplementary Fig. S1. All dielectric and geometric parameters of the hyperbolic media we used here are cited from the actual samples used in published papers or our own experiment, as summarized in Table I. In all these three cases, the hyperbolic polaritons show focusing behavior, and the real part of the field, Re($E_z$), distribution are shown in Fig.

3A-C. FT is performed based on the Re($E_z$) distribution and the results are shown in Fig. 3D-F, visualizing the general dispersions in each case, which are then used to obtain the IFCs through fitting, as shown by the orange dashed curves in Fig. 3D-I. IFC calculated by the TSM in Ref [50] and Eq. 5 in each case are also plotted, as shown by the blue and pink curves in Fig. 3G-3I. The analytical IFC are always showing an agreement with the fitted IFC (orange dash curves) from the dispersion in *k* space. However, the curves from the TSM are more or less deviated from the other two, since the model was obtained with approximations. The deviations of the TSM are greatly affected by the anisotropic level and the thickness of the hyperbolic material. In case of the HMM in near infrared, which has a very strong anisotropy ($|\varepsilon_o/\varepsilon_e| \approx 32$), the TSM is not good enough to describe the hyperbolic character of the IFC, as shown in Fig. 3I.

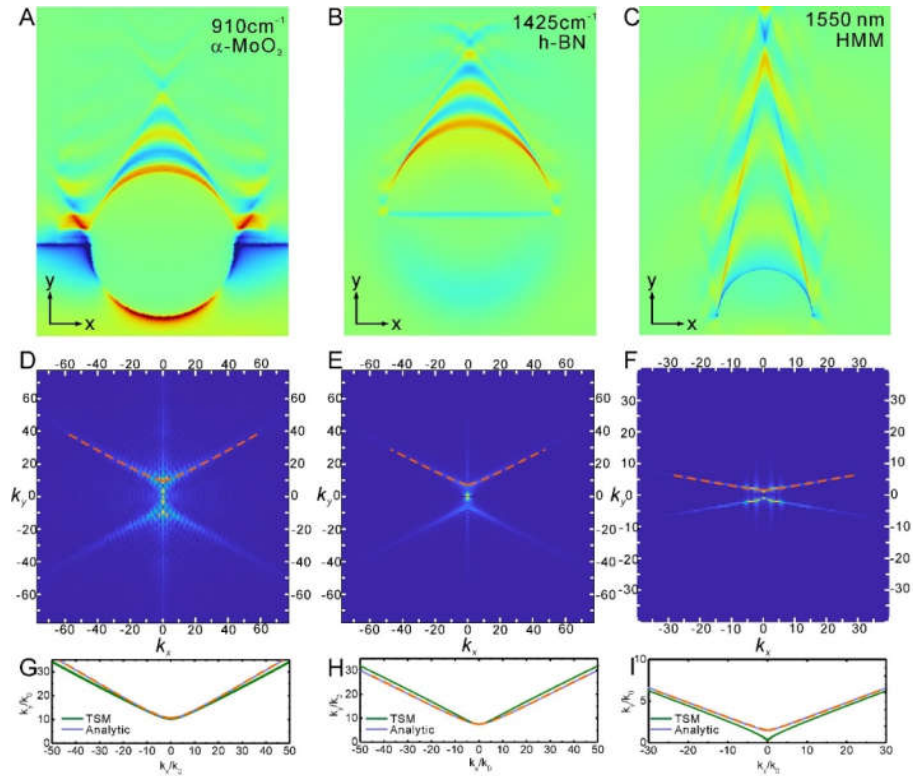

**Fig. 3. Applications of the analytical IFC in commonly used hyperbolic polaritons materials systems.** Numerically simulated Re($E_z$) field distribution of the hyperbolic polaritons along a α-MoO$_3$ flake (**A**), Patterned h-BN (**B**), and HMM film (**B**); (**D-F**) FT images from the Re(Ez) field distribution in panels **a-c**, in three

materials systems; IFCs from TSM, fitted FT images and analytical equation of the α-MoO₃ flake (**G**), Patterned h-BN (**H**), and HMM film (**I**). The orange dash curves indicate the IFCs fitted from the FT images. The blue dash curves indicate the IFCs calculated from Eq. 5. The green dash curves indicate the IFCs calculated from TSM. Details can be found in Methods.

With such an analytical IFC, we can further determine the energy flow and thus trace the wavefront of the hyperbolic polaritons. As we have discussed in Fig. 1F, the direction of the energy flow, namely Poynting vector is normal to the IFC at the angle $\varphi$, with respect to the wave vector at the angle $\theta$:

$$\varphi = \tan^{-1}\left(\frac{dS_y}{dS_x}\right) = \tan^{-1}\left(-\frac{dk_x}{dk_y}\right) \tag{6}$$

where $S_x$ and $S_y$ are the projections of $S$ along $x$ and $y$ axes, respectively, and $-\frac{dk_x}{dk_y}$ can be obtained by taking the derivative of Eq. 5:

$$\frac{dk_x}{dk_y} = \frac{b^2(k_y - c)}{a^2 k_x} \tag{7}$$

Therefore, the energy flow of the hyperbolic polaritons generated at the location ($R\cos\theta$, $R\sin\theta$) at the edge of the round hole can be calculated and the results shown in Fig. S4 implies that all the polaritons at any angle $\theta$ will intersect together at the same location at the y axis, forming a focal spot at $f$, exactly. Since the phase propagation is along $k$, we can project the phase change from the direction $k$ to the direction of $S$ and thus map the wave front of the hyperbolic polaritons in space, which is then demonstrated in the experimental characterization.

In the experiment, the silver grooves based HMM are fabricated by using Focused Ion Beam (FIB) on a 35 nm silver film, with an overall scale of 4×8 μm, as shown in Fig. 4A. At the bottom side, a round hole with 1.2 μm in diameter is milled, as the round source to generate the focused hyperbolic polaritons, which are then characterized by a s-SNOM working at transmission mode. As shown in Fig. 4B, a 1550 nm laser is incident onto the round hole from the substrate side and the E field data (with both amplitude and phase) is collected by the s-SNOM tip on the top surface of the metamaterial. The real part of the E field along z direction and its phase are plotted in Fig. 4C and D, respectively, both of which show the focusing behavior, outlined by the brown dash lines with an angle of 20º, agreeing very well with the value of $2\theta_{min}$ theoretically. Along y direction in Fig. 4C, $l$ indicates the optical distances with the propagation of the hyperbolic polariton along $\theta=90º$. Based on the angular dispersion of $N$ in Eq. 3, the optical distance along any $\theta$ is $N/l$, which is then projected to $\varphi$ direction to locate hyperbolic polariton after the same phase accumulation. This is also the same way we use to derive Eq. 5 in Fig. 1F. By this way, the equi-phase contours are calculated and plotted at the peaks and valleys located along $y$ axis, labeled according to the amplitude data in Fig. 4C. The black and red curves well outline the amplitude profiles in Fig. 4C but also fit the measured equi-phase contour with 0, $\pi$, $2\pi$, and $3\pi$ as shown in Fig. 4D. Meanwhile, we also perform the FT based on the Re($E_z$) in Fig. 4C and the result is shown in Fig. 4E, which again shows excellent agreement between the dispersion behavior in $k$ space and the analytical IFC. The detailed calculation of the equi-phase contour can be found in Supplementary Fig. S4.

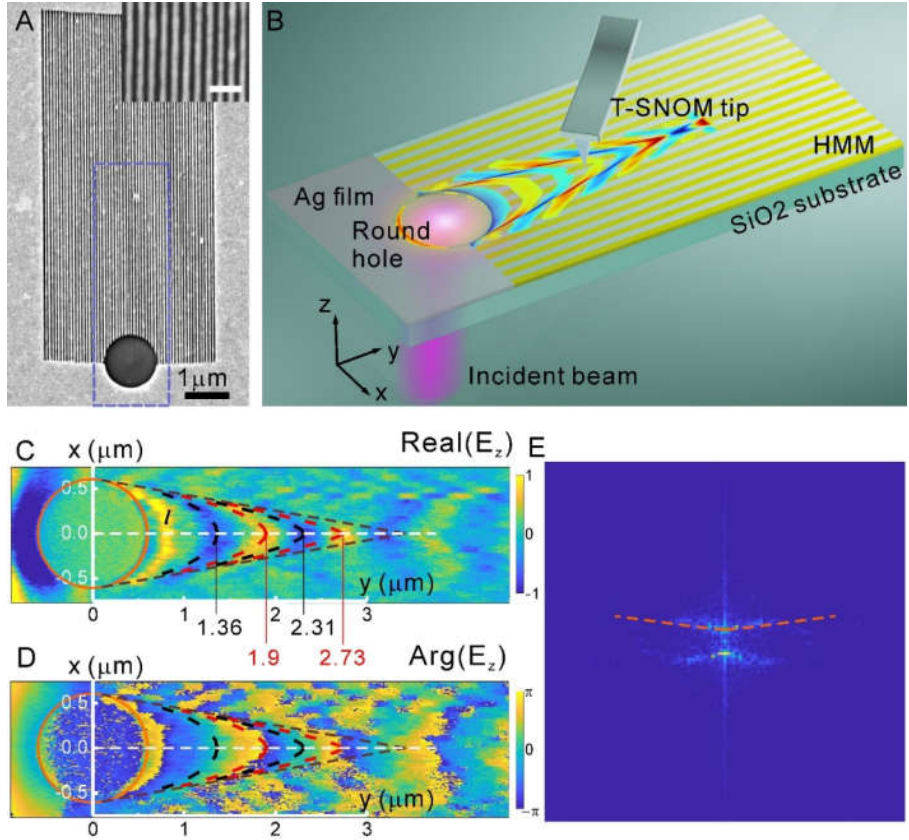

**Fig. 4. Experimental validation of the analytical IFC.** (**A**) SEM image of the fabricated HMM film. The inset shows a magnified view of the silver grooves, whose period is 80 nm with silver stripe width of 40 nm. The scale bar in the inset is 200 nm. (**B**) The schematic of the s-SNOM characterization. (**C**) The Re($E_z$) distribution data obtained by the s-SNOM scanning. The back and red dash curves are the equi-phase contour achieved from the analytical IFC, located at $l$=1.36, 1.9, 2.31 and 2.73 μm, corresponding to the valleys and peaks of the amplitude along $y$ direction. (**D**) The phase distribution of $E_z$ obtained by the s-SNOM scanning. The back and red dash curves achieved from the analytical IFC also fit equi-phase contours measured to be 0, π, 2π, and 3π. (**E**) Analytical IFC plotted in the FT image of the experimental data in C.

**Discussion**

By a thorough study on the focusing behavior of the anisotropic polaritons along a hyperbolic medium film, we obtain a concise analytical form of the IFC, which has been demonstrated to be hyperbolic in the $k$ space, with its central point shifted to the direction with the negative permittivity. The angular dispersion relation of the anisotropic polariton and the normalized refractive index are plotted from this IFC equation, which reveals that the ultra-high confinement in such a system is originated from the ultra-high wave vector when its direction is approaching to the minimum end of the angle band. With such an IFC, we can even precisely obtain the wave propagation analytically. Compared to the calculated IFC from TSM and the fitted IFC from FT of the Re($E_z$), both of which are widely used in literatures, we find that our analytical IFC exhibits excellent agreement with the IFC fitted by FT for all the three materials, and is better than that of the equation from the TSM, which is limited by approximations and shows serious deviations once the hyperbolic material has strong anisotropy. By presenting a clear and intuitive physics image, this work paves the way for developing novel hyperbolic polaritons based optical devices.

**Materials and Methods**

*IFC calculations based on TSM model*

A widely used equation of the hyperbolic polaritons' IFC was derived based on a thin biaxial slab embedded between two semi-infinite media [50] is given by

$$k = \frac{\rho}{d}[\arctan(\frac{\varepsilon_{air}\rho}{\varepsilon_z}) + \arctan(\frac{\varepsilon_s\rho}{\varepsilon_z}) + \pi l], l = 0, 1, 2... \qquad (8)$$

where $k$ is the in-plane wave vector, i.e. $k = \sqrt{k_x^2 + k_y^2}$. The two semi-infinite media in our case are air and glass substrate. The permittivity of air is $\varepsilon_{air}=1$ and the permittivity of the glass is dispersed, which can be found in Table I at certain wavelengths. $d$ is the thickness of biaxial slab, $\rho = i\sqrt{\varepsilon_z/(\varepsilon_x \cos^2\theta + \varepsilon_y \sin^2\theta)}$, and where $\theta$ is the angle between the $x$-axis and $k$, as indicated in Fig. 1.

*Near field characterization*

A s-SNOM (Neaspec GmbH) with transmission mode was utilized to characterize the complex E field information of the hyperbolic polaritons along the HMM film top surface. The transmission mode allows us to manipulate the incident beam and the field collection separately. A Pico second laser (pulse width 8ps) with an operating wavelength of 1550 nm is focused on to the generating source of the round hole from the substrate side. The metal-coated commercial tips (Arrow NCPt-50, NanoWorld) vibrate vertically at a frequency of $\Omega = 270$ kHz is used to collect the scattering field on top of the HMM film surface. Due to the fabrication limit including both the imperfection of the silver film deposition and the FIB milling, the actual loss of silver and the metamaterial would be higher than that from the theoretical effective parameters. Thus, the observed confinement at the focal spot is weakened. However, the loss does not affect the general focusing behavior. According to the amplitude and the phase distribution, we can still determine the wave focused propagation, which are used to test the calculated equi-phase contours. The in norm of the E field measured by s-SNOM is shown in Fig. S5.

**Acknowledgments**

**Funding:**

National Key Research and Development Program of. China (Grants No. 2022YFB3806000)
National Natural Science Foundation of China (Grants No.52332006)

**Supplementary Materials**

**Supplementary Text**

I. Numerical simulation

All the numerical simulations were performed by using *Comsol* Multiphysics 6.0. The materials' parameters of the hyperbolic medium film used in Fig.1D, Fig. 1E, and Fig. 3C were based on a silver/air multilayered structure. The structure is shown in Fig. S1. For a bulky HMM with in-plane anisotropy, the multilayers are standing on the glass substrate. The HMM film can be considered as such a HMM of multilayered structure with a small thickness. Here the multilayers are silver and air with the filling ratio of the silver $p = 0.5$ and the thickness of the hyperbolic medium film $t = 35$ nm. Therefore, the thin multilayered structure is actually a silver nano-grooves' array.

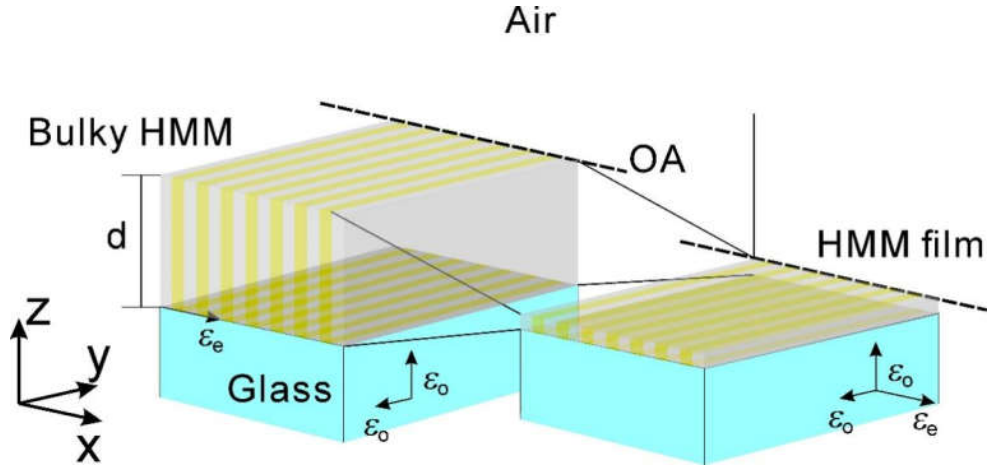

**Fig. S1** The structure of the hyperbolic metamaterial (HMM) film.

According to Maxwell-Garnett theory, the effective permittivity perpendicular to the direction of the multilayers ($\varepsilon_e$) and the permittivity parallel to the direction of the multilayers ($\varepsilon_o$) can be determined by:

$$\varepsilon_o = \varepsilon_{air} p + \varepsilon_{Ag}(1-p)$$
$$\varepsilon_e = \frac{\varepsilon_{air}\varepsilon_{Ag}}{\varepsilon_{Ag} p_{air} + \varepsilon_{air}(1-p)} \quad (s1)$$

where $\varepsilon_{Ag}$ and $\varepsilon_{air}$ are the permittivities of silver and air, respectively. The permittivity of silver is $\varepsilon_{Ag} = -129.165+3.273i$ at $\lambda_0 = 1550$ nm and thus, the effective ordinary permittivity $\varepsilon_o = -64+1.6i$, and extraordinary permittivity $\varepsilon_e = 2$. The substrate under the hyperbolic medium film is glass with permittivity of 2.07. The thickness of the film is 35 nm. In the coordinates of Fig. S1, Fig. 1D, Fig. 1E, and Fig. 3C, $\varepsilon_x = \varepsilon_e$ and $\varepsilon_y = \varepsilon_z = \varepsilon_o$.

Here we would like to have a short discussion on the HMM film and the hyperbolic metasurface structure that are used in Ref. 9 and 10. Although both the two structures are silver grooves, the materials' properties and the polaritons supported at the interfaces are totally different. In our case, the silver grooves are located at the glass substrate and thus, the silver grooves and the air gap compose the HMM in Fig. S1. The surface waves are actually hyperbolic polaritons following the rule of the hyperbolic IFC. Moreover, the polaritons are supported at both the top surface (air/HMM) and the interface between the HMM and the glass substrate. In Ref. 9 and 10, the silver grooves are located at the silver film. According to Ref. 10, at short wavelengths, (e.g. green light or even shorter) the plasmonic modes are tightly confined to the ridges of the grooves, which is qualitatively similar to an array of parallel nanowires that exhibits hyperbolic like dispersion. In such a case, the IFC is actually in a shape of cosine function according to the coupling among the silver nanowires [9, 10]. In the long wavelength limit, (e.g. red light or near infrared), the modes are only weakly confined, and the grooves can be considered a perturbation to a flat surface, just like a bulk silver with groove-shape surface. In this case, the polaritons exist along the ridged surface, which shows elliptical dispersion. In contrast, the HMM shows hyperbolic dispersion in the range from visible to near infrared [38-40]. In our work, the wavelength of the hyperbolic polaritons is 1550 nm, which is too long and can not be focused by the hyperbolic metasurface structure used in Ref. 9 and 10.

## II. IFC in the entire space

Eq. 5 describes the IFC in a half space with $k_y>0$, as expressed by Eq. s2a. In the other half space of $k_y<0$, the dispersion behavior would be the same according to the symmetry of the materials and thus the IFC is also a hyperbola while shifted to - $k_y$ with -$c$, as described by Eq. s2b. Therefore, the total expression of the IFC should include two branches:

$$\frac{(k_y-c)^2}{a^2}-\frac{k_x^2}{b^2}=1, \qquad k_y>0 \quad \text{(s2a)}$$

$$\frac{(k_y+c)^2}{a^2}-\frac{k_x^2}{b^2}=1, \qquad k_y<0 \quad \text{(s2b)}$$

and are plotted in Fig. S2.

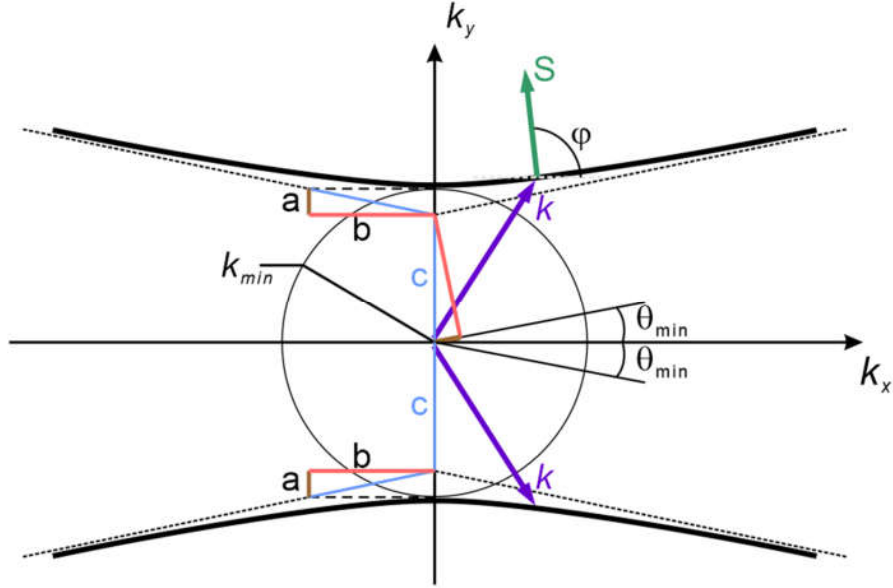

**Fig. S2.** The IFC in the entire space.

### III. Mode with $\theta = \pi/2$

Under $\theta = \pi/2$, the hyperbolic polariton is propagating along $y$-axis. Regarding the TM mode, the E field of the hyperbolic polariton is polarized along the $yz$ plane. Since $\varepsilon_y = \varepsilon_z = \varepsilon_o$, the surface way is decreased to the surface plasmon polariton without any anisotropy. The anisotropic system is also decreased to the metal core waveguide with air cladding on top and glass substrate underneath. In such a case, the wave vector can be obtained by the eigen-equation of the waveguide as shown below:

$$\tanh(\gamma_1 * t) = -\frac{T_2+T_3}{1-T_2 T_3} \qquad \text{(s3)}$$

where $T_2 = \sqrt{\dfrac{\varepsilon_o \gamma_2}{\varepsilon_{air} \gamma_1}}$, $T_3 = \sqrt{\dfrac{\varepsilon_o \gamma_3}{\varepsilon_{glass} \gamma_1}}$, and $\gamma_1 = \sqrt{\beta^2 - k_0^2 \varepsilon_o}$, $\gamma_2 = \sqrt{\beta^2 - k_0^2 \varepsilon_{air}}$,

$\gamma_3 = \sqrt{\beta^2 - k_0^2 \varepsilon_{glass}}$. $\beta$ is the propagation constant and equal to $k_{min}$ with $\theta = \pi/2$ in our system. By using the parameters of the hyperbolic metamaterial film, we can calculate $k_{min}$ by using Eq. s3: $\beta = k_{min} = 1.478 k_0$, which also agrees well with that in the numerical simulation result, as shown in Fig. S3.

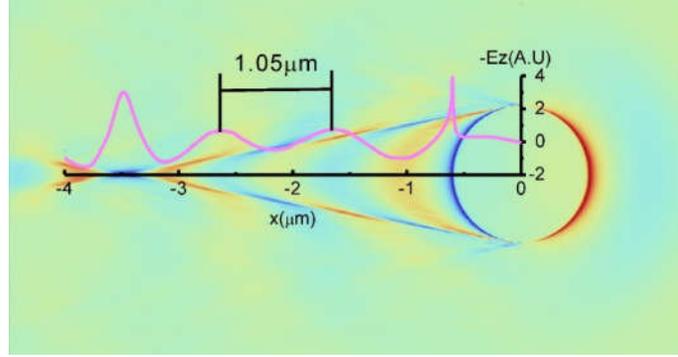

**Fig. S3.** E field distribution of the hyperbolic polariton of $k_{min}$ at $\theta = \pi/2$. The wavelength is measured to be 1.05 μm, which indicates that $k_{min} = 1.48\, k_0$.

## IV. Energy flow and equi-phase contour

The direction of the Poynting vector $S$, which indicates the energy flow is along the normal of the IFC, and thus the refraction angle $\varphi$ can be determined by Eq. 7. Here we calculate the energy flow of the hyperbolic polaritons generated at the location ($R\cos\theta$, $R\sin\theta$) on the round hole and find that all the polaritons at any angle $\theta$ will intersect together at the same spot of the y axis, forming a focal spot at $f = \dfrac{R}{\sin\theta_{min}}$, as shown in Fig. S4.

Furthermore, we can determine the Equi-phase contour by using the analytical IFC. Considering the hyperbolic polariton generated at the edge of the round hole (0, R), both the wave vector k and Poynting vector S are along y direction. If the hyperbolic polariton is generated at ($R\cos\theta$, $R\sin\theta$), the wave vector $k_\theta$ is along $\theta$ but the Poynting vector is along $\varphi$. After a certain propagation along y direction with a length of $l$, there is a phase accumulation of $\Phi$. Along the direction of $\theta$, after the same phase

Φ, the actual propagation length can be calculated by the angular dispersion of the index $N$, as shown in Fig. S4 and the result is $l_\theta = l/N$. Since the energy flow is going along $\varphi$, the actual field with phase Φ, should be located at $(l_{\varphi x}, l_{\varphi y})$ on the corresponding energy flow. By geometry, we can obtain this location as below:

$$l_{\varphi x} = R\cos\theta + \frac{l_\theta}{\cos(\varphi-\theta)}\cos\varphi$$

$$l_{\varphi y} = R\sin\theta + \frac{l_\theta}{\cos(\varphi-\theta)}\sin\varphi$$

(s4)

Therefore, we can realize both the ray tracing and the equi-phase contour mapping through the analytical IFC, as shown in Fig. S4.

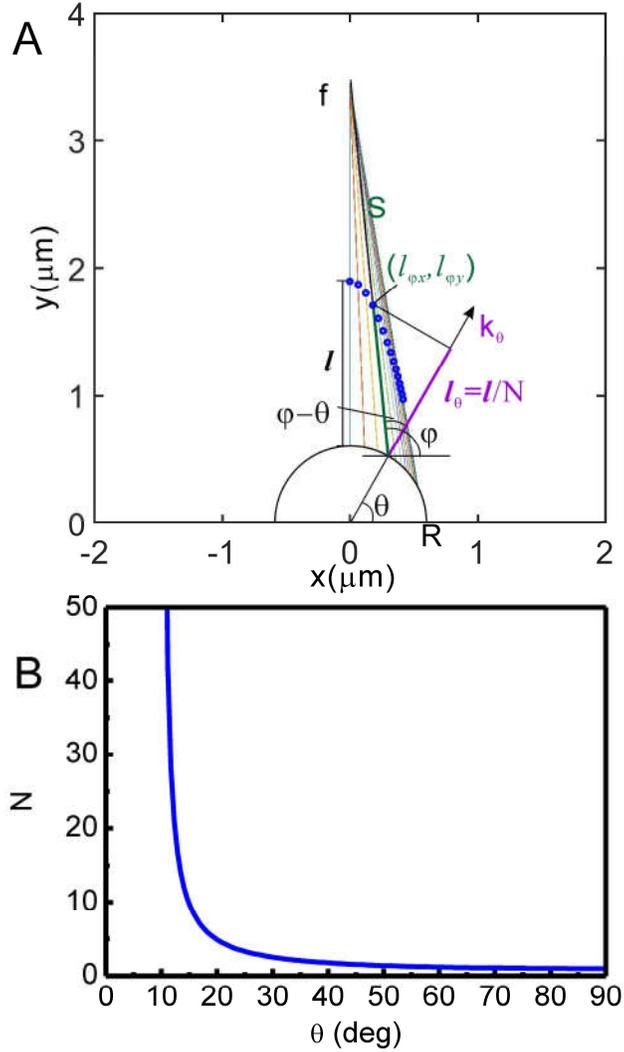

**Fig. S4. Propagation behavior of the hyperbolic polaritons.** (a) Energy flow and the equi-phase contour of the hyperbolic polaritons. All the beams at different angles are all crossed at the same point f. The radius of the round hole that is used to generate the hyperbolic polaritons is R. The green lines shows calculated directions of the Poynting vectors $\varphi$ corresponding to $k_\theta$ at $\theta$. The length $l$ is the propagation distance of the hyperbolic polariton along y direction. Due to the anisotropy, the actual phase propagation along k is $l_\theta=l/N$. By projection to the direction of the Poynting vector, the actual location with the same phase should be ($l_{\varphi x}$, $l_{\varphi y}$) on the corresponding energy flow. (b) The calculated angular dispersion of $N$, calculated based on the Eq. 3 using the parameters from Eq. s3.

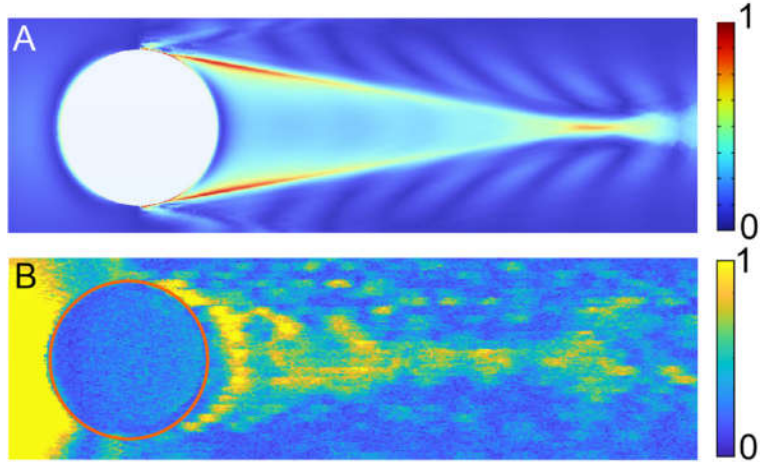

**Fig. S5. The norm of Ez distribution along the hyperbolic metamaterial film.** (a) Simulation result; (b) SNOM result.